\documentclass[%
twocolumn,
superscriptaddress,
nofootinbib,
amsmath,
amssymb,
aps,
prd,
longbibliography
]{revtex4-2}

\usepackage{graphicx}
\usepackage{dcolumn}
\usepackage{bm}

\usepackage{xcolor}

\usepackage{dsfont}
\definecolor{cfull}{HTML}{008080}
\definecolor{clight}{HTML}{84378C}
\definecolor{cstrict}{HTML}{EC008C}
\usepackage{float}
\usepackage{setspace}
\usepackage{graphicx}

\makeatletter
\def\tagform@#1{(\textcolor{magenta}{#1})}
\makeatother
\usepackage[colorlinks=true,linkcolor=magenta,citecolor=magenta,urlcolor=magenta]{hyperref}%

\AtBeginEnvironment{thebibliography}{
  \hypersetup{
    linkcolor=magenta,
    citecolor=magenta,
    urlcolor=magenta
  }
}

\usepackage{lipsum}
\usepackage{xcolor}
\usepackage{textcomp}
\usepackage{amsmath}
\usepackage{amsthm}
\usepackage{amstext}
\usepackage{amssymb}
\usepackage{stmaryrd}
\usepackage{graphicx}
\usepackage{nicefrac}
\usepackage{bbm}
\usepackage{tikz-feynman}
\usepackage{hyperref}

\usetikzlibrary{patterns,decorations.markings}

\tikzset{
cross/.style={fill=white,path picture={\draw[black] (path picture bounding box.south east) -- (path picture bounding box.north west) (path picture bounding box.south west) -- (path picture bounding box.north east);}},
    dressed/.style={fill=white,postaction={pattern=north east lines}},
    momentum/.style 2 args={->,semithick,yshift=5pt,shorten >=5pt,shorten <=5pt},
    loop/.style 2 args={thick,decoration={markings,mark=at position {#1} with {\arrow{>},\node[anchor=\pgfdecoratedangle-90,font=\footnotesize] {$p_{#2}$};}},postaction={decorate}},
    label/.style={thin,gray,shorten <=-1.5ex}
}

\usetikzlibrary{patterns}
\usepackage{xcolor}
\usetikzlibrary{patterns,decorations.markings}

\begin{document}

\title{\textcolor{teal}{ Functional Dimensional Regularization for $O(N)$ Models}}

\author{P. Beretta}
\email{pberetta@fing.edu.uy}
\affiliation{IFFI, Universidad de la Rep\'ublica, J.H.y Reissig 565, 11300 Montevideo, Uruguay}

\author{A. Codello}
\email{alessandro.codello@unive.it}
\affiliation{DSMN, Ca' Foscari University of Venice, Via Torino 155, 30172 - Venice, Italy}
\affiliation{IFFI, Universidad de la Rep\'ublica, J.H.y Reissig 565, 11300 Montevideo, Uruguay}

\begin{abstract}
The novel functional dimensional regularization (FDR) scheme has proven capable of yielding results that are competitive with the state-of-the-art in the computation of critical exponents in $d=3$, while also reproducing those from the $\varepsilon$-expansion for the {\tt Ising} and other universality classes. In this work, we show that this is not a mere coincidence: by applying the scheme to the $O(N)$ universality class, we explicitly derive the flow equations and obtain critical exponents that are comparable to those obtained with higher-order non-perturbative approaches. In this case, FDR retains the features already highlighted in previous works -- namely, its efficiency and rapid convergence.
\end{abstract}

\maketitle



\subsection*{\color{teal}Introduction}\vspace{-0.2cm}

The main idea underlying Functional Dimensional Regularization (FDR) \cite{Beretta:2026zcy,Beretta,Beretta:2026ens} and its applications to critical phenomena is that beta functions/functionals can be defined as a {\it sum over all critical dimensions} of perturbative beta functions/functionals computed in dimensional regularization (DR):
\begin{equation}\label{masterformula}
\beta^{\rm FDR}(d)=\sum_{d_c}^{{\color{white}1}}\mu^{d-d_c}\beta^{\rm DR}(d_c)\,.
\end{equation}
Here $\mu$ is the RG mass scale and $d$ is the dimension of the critical theory under study, which can vary freely. The beta functions/functionals emerging from the sum \eqref{masterformula} are characterized by exponential threshold functions that ensure fast convergence of truncations, while the flow that emerges has all the usual properties of a functional RG \cite{Beretta:2026zcy,Beretta:2026ens}.
The range of the sum over $d_c$ defines which specific variation of the formalism we are using; in equation \eqref{masterformula} we have already restricted the sum to one-loop contributions, i.e. the sum is over even critical dimensions $d_c=2n$ with $n\geq0$.
As explained in \cite{Beretta:2026zcy}, the origin of equation \eqref{masterformula} can be traced to the inclusion/subtraction of all poles appearing in DR, as first hinted by S. Weinberg in \cite{Weinberg} and later applied in various forms to high energy physics \cite{Jack:1990pz,Kaplan:1998tg,Phillips:1998uy,Kluth:2026djl}  and quantum gravity \cite{Kluth:2024lar}.

The scope of this paper is to extend the application of the FDR framework to $O(N)$ symmetry, to acquire a further test of its validity and to compare with other state-of-the-art approaches. 
We will explore the derivative expansion to second order $\partial^2$ (DE2), thus including the wave-function renormalization functionals -- specifically $Z_T$ and $Z_L$ as defined later on -- in the r.h.s. of the RG flow, in addition to the effective potential $U$.
%
%
$O(N)$--symmetry underlies many different universality classes ranging from self-avoiding {\tt SAW} ($N=0$) and loop-erased {\tt LERW} ($N=-2$) random walks, to the classical classes central to statistical mechanics: {\tt Ising} ($N=1$), {\tt XY} ($N=2$) and {\tt Heisenberg} ($N=3$); but also {\tt O(4)} ($N=4$) relevant to high energy physics or the exactly solvable {\tt Spherical-model} in the limit $N\to\infty$.
Owing both to their centrality in the theory of critical phenomena and to the simplicity of the symmetry -- which has only one field invariant entering the Landau--Ginzburg Lagrangian -- basically all theoretical and numerical approaches have something to say on the topic and the related literature is immense (a standard review is \cite{Pelissetto:2000ek}).
Monte Carlo (MC) \cite{Hasenbusch:2010hkk,Clisby:2016} and the (re-summed) $\varepsilon$-expansion ($\varepsilon$E) \cite{Zinn-Justin:2002ecy} have been the traditional benchmark while more recently the Conformal Bootstrap (CB) \cite{Kos:2016ysd,Chester:2020iyt,Dey:2025zgn} has attacked the problem and made significant advances, but for generic values of $N$ the best estimates still come from MC \cite{Hasenbusch:2019jkj,hasenbusch2020monte,Hasenbusch:2021tei} or Non-Perturbative RG (NPRG) \cite{DePolsi:2020pjk,Balog:2019rrg}.
%
%
%
The question is whether the quite remarkable results obtained for the {\tt Ising} class \cite{Beretta:2026zcy,Beretta:2026ens} extend to general values of $N$.
Surprisingly, the results that we report here are competitive with the best NPRG estimates at DE2 level: for $1\leq N\leq 5$ our anomalous dimension is closer to the state-of-the-art values than the NPRG-DE2 one, in some cases approaching DE4 \cite{DePolsi:2020pjk}, while the exponent $\nu$ is somewhat less accurate than at NPRG-DE2. Overall, our estimates compare well with six-loop $\varepsilon$E \cite{Kompaniets:2017yct} and CB estimates \cite{Chester:2019ifh,Kos:2015mba, Kos:2016ysd}, hinting at a deeper validity of \eqref{masterformula} and similar relations at the base of the FDR approach.


\subsection*{\color{teal}Derivative Expansion}\vspace{-0.2cm}

At order DE2 the input action is traditionally chosen as follows \cite{Berges:2000ew}, 
%
%
%
\begin{equation}\label{ansatzUZY}
S=\int_x\biggl\{ 
U(\rho)+\frac{1}{2}Z(\rho)(\partial_\mu \varphi_a)^2+\frac{1}{4}Y(\rho) (\partial_\mu \rho)^2 \biggr\}\,,
\end{equation}
%
with three functions $U,Z,Y$ of the $O(N)$ invariant $\rho=\frac{1}{2}\varphi_{a}\varphi^{a}$.
We find it more convenient to work with the equivalent parametrization in terms of $U,Z_T,Z_L$:
%
\begin{equation}\label{ansatzUTL}
S=\int_x\biggl\{U
+\Big[ Z_T(\delta_{ab}-P_{ab})+Z_L P_{ab}\Big] \frac{1}{2}(\partial_\mu \varphi^a \partial^\mu \varphi^b) \biggr\}\,,
\end{equation}
where we defined the pair of orthogonal projectors: the longitudinal one $P_{ab} = \frac{\phi_a \phi_b}{\phi^2}$ and the transverse one $\delta_{ab}-P_{ab}$.
Clearly the map between the two parametrizations is $Z=Z_T$ and $Y=\frac{Z_L-Z_T}{\rho}$ (and its inverse $Z_T=Z$ and $Z_L=Z+\rho Y$). One advantage of \eqref{ansatzUTL} is the straightforwardness of the identifications in the $N\to 1$ limit where $P_{ab} \to 1$ and $(\delta_{ab}-P_{ab}) \to 0$ so that $Z_{N=1}(\phi) = Z_L(\rho)$.
Another is the simplicity of the normalization. We use the freedom to rescale the field to set $Z_L=1$ at the minimum $\kappa$ of the potential, where we will expand, while at the same point $Z_T=1+\psi_0$, with $\psi_0$ accounting for the difference in normalization of the two wave-functions: $Z_L(\rho)=1+O(\rho-\kappa)$ and $Z_T(\rho)=1+\psi_0+O(\rho-\kappa)$.
From the first condition we will later obtain the anomalous dimension, while $\psi_0$ is a running coupling.
 
%
We start with a pedagogical derivation of the beta functional of the potential  to explain how to use \eqref{masterformula} in practice.
The Hessian of the action \eqref{ansatzUTL}, at constant fields, reads
\begin{equation}
\mathbb{S}^{(2)}(q^2)= \big[\omega_T+q^2 Z_T\big](\mathbbm{1}-\mathbb{P})+\big[\omega_L+q^2 Z_L\big]\mathbb{P}\,,
\end{equation}
where $\omega_T=U'$, $\omega_L=U'+2\rho U''$, $Z_T=Z$ and $Z_L=Z+\rho Y$.
The one-loop effective action can be computed directly from it in a standard way
\begin{equation}\label{trlog}
\Gamma_1= \frac{1}{2} {\rm Tr} \log \mathbb{S}^{(2)} = 
-\frac{\Omega}{2}\int_{0}^{\infty}\frac{{\rm d}s}{s}\int_q{\rm tr}\, e^{-s\,\mathbb{S}^{(2)}(q^2)} \, .
\end{equation}
The index trace can be evaluated using the properties of the projectors introduced earlier, leading to
\begin{equation}
{\rm tr}\, e^{-s\,\mathbb{S}^{(2)}(q^2)} = (N-1)e^{-s[q^2 Z_T+\omega_T]}+e^{-s[q^2 Z_L+\omega_L]} \, .
\end{equation}
The momentum and proper-time integrals can now be easily performed for general dimension, giving (per unit volume)
%
%
\begin{equation*}
\Gamma_1= -\frac{\Gamma\big(-\frac{d}{2}\big)}{2(4\pi)^{\frac{d}{2}}} 
\biggl\{(N-1) \left(\frac{\omega_T}{Z_T}\right)^{\frac{d}{2}}
+\left(\frac{\omega_L}{Z_L}\right)^{\frac{d}{2}} \biggr\}\,.
\end{equation*}
The Gamma function has poles whenever $d=2n\equiv d_c$, with $n$ a non-negative integer. 
Our FDR beta functions \eqref{masterformula} are built out of the DR beta functions at each $d_c$, which requires identifying the $\frac{1}{\varepsilon}$-poles in the complex $d$-plane. We just need the relation
$\Gamma\big(-\frac{d}{2}\big) = \frac{1}{\varepsilon}(-1)^{d_c/2}\frac{2}{(d_c/2)!} + \dots$, valid for $d=d_c-\varepsilon$,
and to remember that the DR beta functional at $d_c$ is minus the respective residue
\begin{equation*}
\beta^{\rm DR}_U(d_c) =  \frac{(-1)^{\frac{d_c}{2}}}{(4\pi)^\frac{d_c}{2}\big(\frac{d_c}{2}\big)!}
\biggl\{(N-1)\left(\frac{\omega_T}{Z_T}\right)^{\!\frac{d_c}{2}} +\left(\frac{\omega_L}{Z_L}\right)^{\!\frac{d_c}{2}} \biggr\}\, .
\end{equation*}
Finally we apply the master formula \eqref{masterformula} with sum over all critical one-loop dimensions $d_c=0,2,4,6,8,...$,
\begin{equation}
\beta_U(d)=\sum_{d_c}\mu^{d-d_c}  \beta^{\rm DR}_U(d_c) \,.
\end{equation}
The series sums to an exponential and returns the FDR--DE2 beta functional for the $O(N)$ potential
\begin{equation}\label{FDR-BetaU}
\beta_U = \mu^d \,e^{-\tfrac{U'+2\rho U''}{4\pi\mu^2  Z_L}} + \mu^d \,(N-1)\, e^{-\tfrac{U'}{4\pi\mu^2 Z_T}}\, .
\end{equation}
This expression has a longitudinal ($L$) and a transverse ($T$) contribution (proportional to $N-1$), as is the case for functional flows with this symmetry.
In the limit $N\to 1$ only the former survives and we recover the {\tt Ising} flow at order FDR-DE2 \cite{Beretta:2026ens}
\begin{equation}
\beta_V(\phi) = \mu^d e^{ -\tfrac{V''}{4\pi\mu^2 Z}}\,,
\end{equation}
with $Z(\phi) = Z_L(\rho)$ and $V''(\phi) = U'(\rho)+2\rho U''(\rho)$.
In the opposite limit $N\to\infty$ only the transverse part survives (a limit that we will not explore here).
As discussed in \cite{Beretta:2026zcy}, the FDR beta functional for the potential is equivalent to the one of the proper-time RG (PTRG) in the limit $m\to \infty$ of the PTRG regulator parameter $m$. This is also true in the $O(N)$ extension, and our beta functional \eqref{FDR-BetaU} agrees with \cite{Mazza:2001bp} after a field redefinition.
It is at order DE2 that they start to differ (see \cite{Bonanno:2000yp} for the PTRG at this order in the {\tt Ising} case). 

The beta functionals of the wave-function renormalizations, $Z_L$ and $Z_T$, can be computed along similar lines -- first determining $\beta^{\rm DR}_{Z_T}(d_c)$ and $\beta^{\rm DR}_{Z_L}(d_c)$ and then summing over $d_c$ as we did for the potential; full details of the derivation can be found in \cite{Beretta}. The final result can be cast in the following form:
\begin{widetext}
\begin{eqnarray}
\beta_{Z_L} & =&\left\{-\frac{\mu^{d-2}}{4\pi}\frac{Z_{L} (Z_{L}'+2\rho Z_{L}'')-3\rho(Z_{L}')^{2}}{ Z_{L}^{2}}
+\frac{\mu^{d-4}}{(4\pi)^{2}}\frac{2\rho Z_{L}'(-U'Z_{L}'+U''(3Z_{L}-2\rho Z_{L}')+2\rho U'''Z_{L})}{Z_{L}^{3}}
\right.  \nonumber\\
 && \qquad\left. -\frac{\mu^{d-6}}{(4\pi)^{3}}\frac{\rho(U'Z_{L}'-U''(3Z_{L}-2\rho Z_{L}')-2\rho U'''Z_{L})^{2}}{3Z_{L}^{4}}\right\}e^{-\tfrac{U'+2\rho U''}{4\pi\mu^{2}Z_{L}}}  \nonumber\\
 && -(N-1)\left\{ \frac{\mu^{d-6}}{(4\pi)^{3}}\frac{\rho(U''Z_{T}-U'Z_{T}')^{2}}{3Z_{T}^{4}}+\frac{\mu^{d-4}}{(4\pi)^{2}}\frac{2(U''Z_{T}-U'Z_{T}')(Z_{T}-Z_{L}+\rho Z_{T}')}{Z_{T}^{3}}\right. \nonumber\\
 && \qquad\qquad\qquad\left.+\frac{\mu^{d-2}}{4\pi}\frac{\rho Z_{T}(Z_{L}'+2Z_{T}')+Z_{T}^{2}-Z_{L}(Z_{T}+2\rho Z_{T}')+\rho^{2}(Z_{T}')^{2}}{\rho Z_{T}^{2}}\right\} e^{-\tfrac{U'}{4\pi\mu^{2}Z_{T}}}\,, \label{betaL}\\[2mm]
\beta_{Z_T} & =&\left\{\mu^d\frac{2(Z_{T}+\rho Z_{T}')^{2}}{\rho(Z_{T}-Z_{L})U'+2\rho^{2}Z_{T}U''}-\frac{\mu^{d-2}}{4\pi}\frac{(N-1)\rho Z_{T}'+Z_{T}}{\rho Z_{T}}\right\}e^{-\tfrac{U'}{4\pi\mu^{2}Z_{T}}} \nonumber \\
 && -\left\{\mu^{d}\frac{2(Z_{T}+\rho Z_{T}')^{2}}{\rho(Z_{T}-Z_{L})U'+2\rho^{2}Z_{T}U''}+\frac{\mu^{d-2}}{4\pi}\frac{Z_{T}+5\rho Z_{T}'+2\rho^{2}Z_{T}''}{\rho Z_{L}}\right\} e^{-\tfrac{U'+2\rho U''}{4\pi\mu^{2}Z_{L}}}\,. \label{betaT}
\end{eqnarray}
\end{widetext}
The formula for $\beta^{\rm DR}_{Z_L}(d_c)$ appears also in the Appendix of \cite{Baldazzi:2020vxk} and it can be checked that \eqref{betaL} can be obtained from this expression via our main relation \eqref{masterformula}.
Even if \eqref{betaT} and \eqref{betaL} cannot be called ``simple'', they are still much more concise than the corresponding expressions in the NPRG formalism (see for example \cite{VonGersdorff:2000kp,Chlebicki:2020pvo}).
As an important check, in the $N\to 1$ limit these expressions reduce to those presented in \cite{Beretta:2026ens}.
%

\subsection*{\color{teal}LPA Analysis}\vspace{-0.2cm}

In terms of dimensionless variables -- $u(\rho), z_T(\rho)$ and $z_L(\rho)$ -- the flow equation for the potential takes the following form\footnote{We will use the symbol $\rho$ for both the dimensionful and dimensionless cases; no confusion should arise.}
\begin{equation}\label{FDR-LPA}
\beta_u=-d u+(d-2+\eta )\rho u'+(N-1)\,e^{-\frac{u'}{4\pi z_T}}+e^{-\frac{u'+2\rho u''}{4\pi z_L}}\,.
\end{equation}
In this section we start with the local potential approximation (FDR-LPA), where we set $z_T=z_L=1$ so that only the dimensionless potential $u$ flows.
Clearly $\eta =0$ in this setting.
Although \eqref{FDR-LPA} is -- in principle -- valid in any dimension, from now on we will fix $d=3$.
 Due to the rapid convergence offered by the exponential threshold function in \eqref{FDR-LPA} it is possible to study scaling solutions $\beta_u=0$ using a simple polynomial expansion around the minimum $\kappa$ of the potential 
\begin{equation}\label{PolyExp}
u(\rho)= 
\tfrac{u_2}{2!} (\rho-\kappa)^2 + \tfrac{u_3}{3!}  (\rho-\kappa)^3  + \tfrac{u_4}{4!}  (\rho-\kappa)^4 + ...
\end{equation}
Here $u_{k}$ are the running coupling constants and -- as usual -- their beta functions $\beta_{k}$ are obtained upon insertion of \eqref{PolyExp} into both sides of \eqref{FDR-LPA} followed by a series expansion to compare coefficients.
We then truncate the system of beta functions to a subset of $n_u$ couplings and study the Wilson-Fisher fixed point in successive approximations. The computation of the stability matrix then leads to the RG eigenvalues $\theta_k$, from which we extract $\nu = -1/\theta_2$ and $\omega = \theta_3$.

Fast convergence is observed as the number of couplings $n_u$ is increased (it is enough to use $n_u\sim 10$) and one can appreciate the quality of the estimates which are quite close to the state-of-the-art results for all values considered.
Results for $N=0,1,2,3,4,5,10$ are reported in Table \ref{Results} and compared with the literature.
We notice the quite remarkable fact that for $1\leq N\leq 5$ our FDR-LPA estimates of the critical exponent $\nu$ agree with the state-of-the-art results to two significant digits, and for $N=0$ and $N=10$ they differ from them by less than $0.01$ (this is not the case with the standard NPRG-LPA). The situation for the exponent $\omega$ is poorer: for $N\leq3$ and $N=10$ our estimates are still better than the corresponding NPRG-LPA ones \cite{DePolsi:2020pjk,Litim:2002cf}, while for $N=4,5$ they are slightly worse.
We have also computed sub-leading exponents (cf.\ \cite{Litim:2002cf}); for completeness, they can be found in \cite{Beretta}.

As explained in \cite{Beretta:2026zcy}, the FDR-LPA coincides exactly with the PTRG-LPA in the limit $m\to \infty$ (while crucially differing from order DE2 onwards). Our values agree with those first computed in this context by \cite{Mazza:2001bp} and later by \cite{Litim:2001hk}.
As already commented, these LPA values are very good, and we think that FDR -- being the functional incarnation of DR -- ``explains'' the success of PTRG estimates, even though the latter flow is not exact. See also \cite{Litim:2001dt} for a study of $\nu(N)$ and $\omega(N)$ in the NPRG-LPA for various cutoffs and for comparison with PTRG results.
%

\subsection*{\color{teal}DE2 Analysis}\vspace{-0.2cm}

\begin{figure}[t]
\centering
\includegraphics[width=\columnwidth]{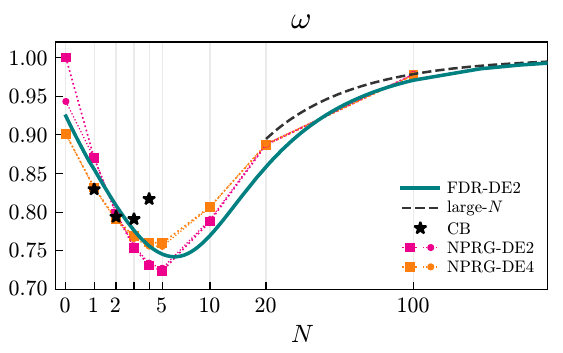}
\caption{Critical exponent $\omega$ as a function of $N$ in $d=3$: FDR-DE2 with $n_u=11$ (teal), compared with NPRG-DE2 (magenta) and NPRG-DE4 (orange) \cite{DePolsi:2020pjk} (squares: final results, dots: raw data), the large-$N$ expansion (dark grey, dashed) and CB estimates (stars). The horizontal axis is linear in $\log(N+3)$.}
\label{figomega}
\end{figure}

\begin{figure}[t]
\centering
\includegraphics[width=\columnwidth]{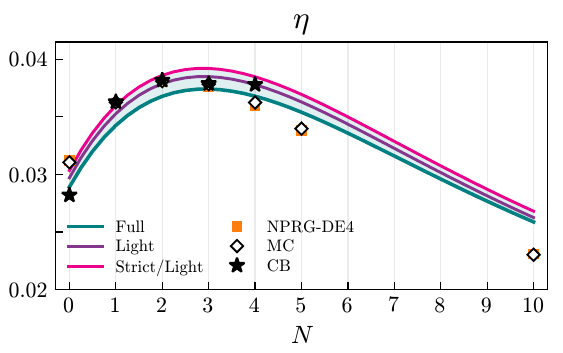}
\caption{Anomalous dimension $\eta$ as a function of $N$ in $d=3$ obtained with the {\color{cfull}Full} (teal), {\color{clight}Light} (purple) and {\color{cstrict}Strict/Light} (magenta) FDR-DE2 flows defined in Appendix~\ref{appC} (expansion around $\rho=0$ with $n_u=11$); the shaded band spans the three results. Squares: NPRG-DE4 \cite{DePolsi:2020pjk}; diamonds: MC \cite{Clisby:2016,Clisby:2017,Hasenbusch:2010hkk,Hasenbusch:2019jkj,hasenbusch2020monte,Hasenbusch:2021tei}; stars: CB \cite{Shimada:2015gda,Kos:2016ysd,Chester:2019ifh,Chester:2020iyt,Kos:2015mba}.}
\label{figband}
\end{figure}

We are now ready to present the full DE2 analysis.
In addition to \eqref{FDR-LPA} we consider the dimensionless version of equations \eqref{betaT} and \eqref{betaL}
which read:
\begin{eqnarray}\label{betaDE2dimless}
\beta_{z_L}&=&\eta z_L+(d-2+\eta)\rho z_L' + \mu^{\eta}\beta_{Z_L} \nonumber\\
\beta_{z_T}&=&\eta z_T+(d-2+\eta)\rho z_T' + \mu^{\eta}\beta_{Z_T}\,.
\end{eqnarray}
At DE2 we expand all three functions around the minimum $\kappa$ of the potential:
\begin{eqnarray}\label{truncation}
u(\rho)&=&\tfrac{u_2}{2!}(\rho-\kappa)^2+\tfrac{u_3}{3!}(\rho-\kappa)^3+\dots\,,\nonumber\\
z_L(\rho)&=&1+\zeta_1(\rho-\kappa)+\tfrac{\zeta_2}{2!}(\rho-\kappa)^2+\dots\,,\nonumber\\
z_T(\rho)&=&1+\psi_0+\psi_1(\rho-\kappa)+\tfrac{\psi_2}{2!}(\rho-\kappa)^2+\dots\,,
\end{eqnarray}
which introduces, besides $\kappa$ and the potential couplings $u_k$, two sets of wave-function-renormalization couplings, $\zeta_{k}$ and $\psi_{k}$.
We keep $n_u$ potential couplings, $\kappa,u_2,\dots,u_{n_u}$, and $n_z=n_u+2$ couplings in $z_L$, $\zeta_1,\dots,\zeta_{n_z}$, and in $z_T$, $\psi_1,\dots,\psi_{n_z}$, besides $\psi_0$.
Since the expansion point runs, the flow of the couplings follows from \eqref{FDR-LPA} and \eqref{betaDE2dimless} evaluated at $\rho=\kappa$:
\begin{equation}\label{kappaflow}
\partial_t\kappa=-\tfrac{\beta_u'(\kappa)}{u_2}\,,\qquad
\partial_t u_k=\beta_u^{(k)}(\kappa)+u_{k+1}\,\partial_t\kappa\,,
\end{equation}
and analogously $\partial_t\zeta_k=\beta_{z_L}^{(k)}(\kappa)+\zeta_{k+1}\partial_t\kappa$ and $\partial_t\psi_k=\beta_{z_T}^{(k)}(\kappa)+\psi_{k+1}\partial_t\kappa$ for $k\geq0$.
\begin{figure}[t]
\centering
\includegraphics[width=\columnwidth]{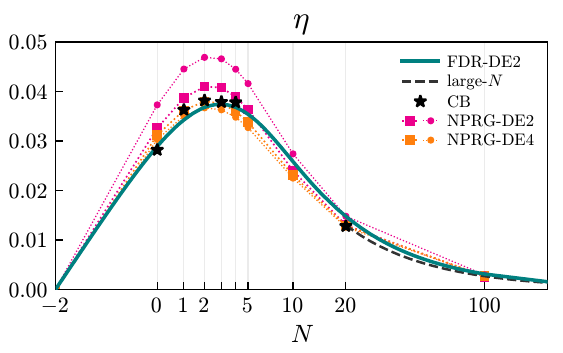}
\caption{Critical exponent $\eta$ as a function of $N$ in $d=3$: FDR-DE2 with $n_u=11$ (teal), compared with NPRG-DE2 (magenta) and NPRG-DE4 (orange) \cite{DePolsi:2020pjk} (squares: final results, dots: raw data), the large-$N$ expansion (dark grey, dashed) and CB estimates (stars). The horizontal axis is linear in $\log(N+3)$.}
\label{figeta}
\end{figure}
\begin{figure}[t]
\centering
\includegraphics[width=\columnwidth]{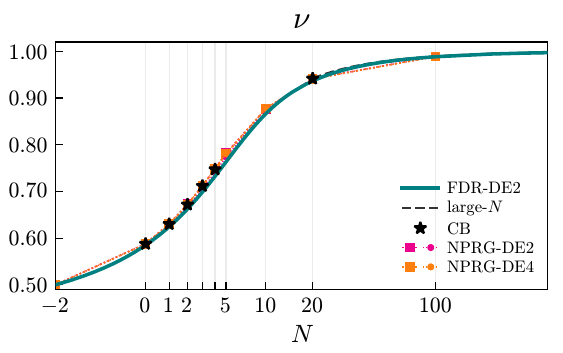}
\caption{Critical exponent $\nu$ as a function of $N$ in $d=3$: FDR-DE2 with $n_u=11$ (teal), compared with NPRG-DE2 (magenta) and NPRG-DE4 (orange) \cite{DePolsi:2020pjk} (squares: final results, dots: raw data), the large-$N$ expansion (dark grey, dashed) and CB estimates (stars). The horizontal axis is linear in $\log(N+3)$.}
\label{fignu}
\end{figure}
The anomalous dimension follows from requiring that the normalization $z_L(\kappa)=1$ is preserved by the flow, $\partial_t\zeta_0=\beta_{z_L}(\kappa)+\zeta_1\partial_t\kappa=0$, which is a linear equation for $\eta$.
The constant $\psi_0$, instead, is a genuine running coupling, $\partial_t\psi_0=\beta_{z_T}(\kappa)+\psi_1\partial_t\kappa$, with a non-vanishing fixed-point value (e.g.\ $\psi_0^*\simeq-0.038$ for $N=1$ and $\psi_0^*\simeq-0.054$ for $N=3$); setting $\psi_0=0$ is not consistent with the flow.
Alternatively, one can normalize the transverse wave-function, $z_T(\kappa)=1$, letting $z_L(\kappa)$ run and extracting $\eta$ from $\beta_{z_T}$. The two expressions for $\eta$ so determined are different functions of the couplings but, since the flow is invariant under rescalings of the field, they crucially agree at the fixed point (in our numerics to all computed digits).
The polynomial expansion around $\kappa$ does not enforce the regularity condition $Z_T(0)=Z_L(0)$ (i.e.\ a $Y$ regular at the origin), which the fixed point satisfies automatically. As a consequence, the stability matrix contains an additional eigenvalue whose eigenvector violates this condition and which has to be discarded; for $N=0$ it is negative, $\theta\simeq-0.35$, and should not be mistaken for $-1/\nu$. We checked that expanding instead around $\rho=0$, where $Z_T(0)=Z_L(0)$ holds identically and $\eta$ takes the simple form \eqref{eta_rho0} given below, leads to the same critical exponents.
As in the LPA analysis we compute critical exponents via the stability matrix after determining the coordinates of the Wilson-Fisher fixed point.
Truncations converge rapidly -- to obtain four significant digits for the critical exponents we need just $n_u\sim 15$ on a desktop computer. We also remark that we do not need any optimization, since {\it there is no regulator} and our calculation yields a unique result for each critical quantity under study.
The DE2 beta functionals contain terms that are formally of higher order in the wave-function renormalizations, whose treatment is not fixed by the derivative expansion. In Fig.~\ref{figband} we compare the anomalous dimension obtained with the full flow with two simplified versions of it, Light and Strict/Light, defined in Appendix~\ref{appC}. Since the three versions are equally consistent at order DE2, the band they span provides a proxy for the truncation error of the DE2 estimate of $\eta$. Its width is about $5\%$ of $\eta$ for $N\lesssim5$ (between $0.0014$ and $0.0018$) and decreases to about $4\%$ at $N=10$. For $N=2,3$ the band contains both the CB and the MC values, and for $0\leq N\leq5$ the best estimates lie within about one band width from it. At $N=10$, instead, the FDR estimate exceeds the MC value by about three band widths, signalling that the DE2 error grows with $N$ faster than the spread.
Our results for the specific values $N=0,1,2,3,4,5,10$ are shown in Table \ref{Results} and compare well with state-of-the-art NPRG, CB and $\varepsilon$E estimates\footnote{MC and CB give more precise determinations of the critical exponents, but at a much larger computational cost: several years of CPU time for MC and about $10^2$ years for CB, as estimated in \cite{DePolsi:2020pjk}. By comparison, with a simple Python implementation of the FDR-DE2 flow, a fixed point and its stability matrix at $n_u=12$ are obtained in about ten seconds on a single CPU core.}. 
More generally, the low computational cost of our approach and its fast convergence allow for a dense sampling in $N$ between $N=-2$ and $N=1000$ which furnishes the critical exponents {\it functions} $\eta(N)$, $\nu(N)$ and $\omega(N)$ with unprecedented precision, 
as displayed in Figures \ref{figeta}, \ref{fignu} and \ref{figomega},
where comparison is made with NPRG-DE4 and CB results.
Finally, for $N\to\infty$ the FDR exponents approach the exact values $\nu=1$, $\omega=1$ and $\eta=0$, which FDR reproduces exactly at leading order, while the $1/N$ corrections are captured only approximately at order DE2, as discussed below.
\renewcommand{\arraystretch}{0.95}
\begin{table*}[t]
\centering
\begin{tabular}{
c| 
c|
l l |                               
l l l                               
c| 
lc| 
l c| 
lc
}
\hline\hline
 &  & \multicolumn{2}{l|}{\bf FDR} & \multicolumn{3}{l}{\bf NPRG} &  & \multicolumn{2}{l|}{\bf CB} &  \multicolumn{2}{l|}{\bf $\varepsilon$E} &\multicolumn{2}{l}{\bf MC} \\
\hline
UC & Exp. & LPA & $O(\partial^{2})$ & LPA & $O(\partial^{2})$ & $O(\partial^{4})$ & Ref. &  & Ref. &  & Ref.& & Ref.\\
\hline
& $\eta$ & 0 & {\bf 0.02891} & 0 & 0.0326 & {\bf 0.0312} & & {\bf 0.0282} & & {\bf 0.0310}(7) & & {\bf 0.031043}(3) & \\
$N=0$ & $\nu$ & 0.5828 & 0.5851 & 0.5925 & 0.5879 & 0.5876 & \cite{DePolsi:2020pjk} & 0.5876 & \cite{Shimada:2015gda} & 0.5874(3) & \cite{Kompaniets:2017yct} & 0.5875970(4) & \cite{Clisby:2016,Clisby:2017}\\
& $\omega$ & 0.7819 & 0.925 & 0.66 & 1.00 & 0.901 & & -- & & 0.841(13) & & 0.899(14) & \\
\hline
& $\eta$ & 0 & {\bf 0.03426} & 0 & 0.0387 & {\bf 0.0362} & & {\bf 0.03630} & \cite{Kos:2016ysd} & {\bf 0.0362}(6) & & {\bf 0.03627}(10) & \\
$N=1$ & $\nu$ & 0.6260 & 0.6244 & 0.6496 & 0.6308 & 0.62989 & \cite{DePolsi:2020pjk} & 0.6300 & \cite{Kos:2016ysd} & 0.6292(5) & \cite{Kompaniets:2017yct} & 0.63002(10) & \cite{Hasenbusch:2010hkk}\\
& $\omega$ & 0.7622 & 0.8555 & 0.654 & 0.870 & 0.832 & & 0.8295 & \cite{Reehorst:2021hmp} & 0.820(7) & & 0.832(6) & \\
\hline
& $\eta$ & 0 & {\bf 0.03679} & 0 & 0.0410 & {\bf 0.0380}(13) & & {\bf 0.03818} & & {\bf 0.0380}(6) & & {\bf 0.03810}(8) & \\
$N=2$ & $\nu$ & 0.6690 & 0.6624 & 0.7090 & 0.6725 & 0.6716(6) & \cite{DePolsi:2020pjk} & 0.6718 & \cite{Chester:2019ifh} & 0.6690(10) & \cite{Kompaniets:2017yct} & 0.67169(7) & \cite{Hasenbusch:2019jkj}\\
& $\omega$ & 0.7457 & 0.8088 & 0.672 & 0.798 & 0.791(8) & & 0.794 & & 0.804(3) & & 0.789(4) & \\
\hline
& $\eta$ & 0 & {\bf 0.03742} & 0 & 0.0408 & {\bf 0.0376}(13) & & {\bf 0.03788} & \cite{Chester:2020iyt} & {\bf 0.0378}(5) & & {\bf 0.03784}(5) & \\
$N=3$ & $\nu$ & 0.7103 & 0.6988 & 0.7620 & 0.7125 & 0.7114(9) & \cite{DePolsi:2020pjk} & 0.7117 & \cite{Chester:2020iyt} & 0.7059(20) & \cite{Kompaniets:2017yct} & 0.71164(10) & \cite{hasenbusch2020monte}\\
& $\omega$ & 0.7346 & 0.7767 & 0.702 & 0.754 & 0.769(11) & & 0.791 & \cite{CastedoEcheverri:2016fxt} & 0.795(7) & & 0.759(2) & \\
\hline
& $\eta$ & 0 & {\bf 0.03680} & 0 & 0.0389 & {\bf 0.0360} & & {\bf 0.0378} & \cite{Kos:2015mba} & {\bf 0.0366}(4) & & {\bf 0.03624}(8) & \\
$N=4$ & $\nu$ & 0.7485 & 0.7326 & 0.805 & 0.749 & 0.7478 & \cite{DePolsi:2020pjk} & 0.7472 & \cite{Kos:2015mba} & 0.7397(35) & \cite{Kompaniets:2017yct} & 0.74817(20) & \cite{Hasenbusch:2021tei}\\
& $\omega$ & 0.7299 & 0.7562 & 0.737 & 0.731 & 0.761 & & 0.817 & \cite{CastedoEcheverri:2016fxt} & 0.794(9) & & 0.755(5) & \\
\hline
& $\eta$ & 0 & {\bf 0.03540} & 0 & 0.0364 & {\bf 0.0338} & & -- & & 0.034$^\dagger$ & & {\bf 0.03397}(9) & \\
$N=5$ & $\nu$ & 0.7825 & 0.7634 & 0.839 & 0.782 & 0.7797 & \cite{DePolsi:2020pjk} & -- & & 0.766$^\dagger$ & \cite{Antonenko:1995plu} & 0.7802(6) & \cite{Hasenbusch:2021tei}\\
& $\omega$ & 0.7317 & 0.7453 & 0.770 & 0.724 & 0.760 & & -- & & -- & & 0.754(7) & \\
\hline
& $\eta$ & 0 & {\bf 0.02585} & 0 & 0.0240 & {\bf 0.0231} & & -- & & 0.024$^\dagger$ & & {\bf 0.02302}(12) & \\
$N=10$ & $\nu$ & 0.8868 & 0.8664 & 0.919 & 0.877 & 0.8776 & \cite{DePolsi:2020pjk} & -- & & 0.859$^\dagger$ & \cite{Antonenko:1995plu} & 0.8797(9) & \cite{Hasenbusch:2021tei}\\
& $\omega$ & 0.7906 & 0.7697 & 0.874 & 0.788 & 0.807 & & -- & & -- & & 0.816(16) & \\
\hline\hline
\end{tabular}
\caption{Critical exponents of the $O(N)$ universality classes in $d=3$ for $N=0,1,2,3,4,5,10$ obtained with different methods and approximation orders: FDR (this work), NPRG \cite{DePolsi:2020pjk} (final results in the nomenclature of that reference), conformal bootstrap (CB), six-loop $\varepsilon$-expansion ($\varepsilon$E) and Monte Carlo (MC). For $N=5$ and $10$, the values marked with $^\dagger$ are from six-loop perturbation theory directly in $d=3$ \cite{Antonenko:1995plu}. The CB values of $\omega$ for $N=3,4$ are from the non-rigorous effective bootstrap \cite{CastedoEcheverri:2016fxt}.}
\label{Results}
\end{table*}
\renewcommand{\arraystretch}{1}

\subsection*{\color{teal}Recovering the $\varepsilon$-expansion}\vspace{-0.2cm}

As a consistency check of the derivative expansion at $O(\partial^2)$, we now show that the FDR flow reproduces the two-loop result for the anomalous dimension of the $O(N)$ model close to four dimensions. Even though the master formula \eqref{masterformula} only involves one-loop DR beta functionals, the anomalous dimension is correctly obtained at order $\varepsilon^2$: the field dependence of the wave-function renormalizations is generated at order $\lambda_4^2$ by the polarization diagram, and it feeds back into $\eta$ through Eq.~\eqref{eta_rho0} below. This mechanism is the same as in the {\tt Ising} case \cite{Beretta:2026ens}.

Close to four dimensions it is convenient to expand around $\rho=0$, $u=\sum_{k\geq1}\frac{\lambda_{2k}}{k!}\rho^k$, $z_L=1+\zeta_2\rho+\dots$ and $z_T=1+\psi_2\rho+\dots$, where regularity of $Y$ implies $Z_T(0)=Z_L(0)$, so that both wave-functions can be normalized to one. At $\rho=0$ the beta functionals \eqref{betaT} and \eqref{betaL} coincide identically, $\beta_{Z_T}(0)=\beta_{Z_L}(0)$, so this normalization is preserved by the flow, and the $\rho\to0$ limit of $\beta_{z_L}$ (in which the transverse term proportional to $1/\rho$ has a finite limit) gives the exact expression
\begin{equation}
\eta
=
\tfrac{(N-1)\psi_2+\zeta_2}{4\pi}
e^{-\frac{\lambda_2}{4\pi}}\,,
\label{eta_rho0}
\end{equation}
with no need for further approximations.
Close to $d=4$ the Wilson--Fisher fixed point is perturbative, with $\lambda_2,\lambda_4=O(\varepsilon)$ and $\psi_2,\zeta_2=O(\varepsilon^2)$, so that $\eta=[(N-1)\psi_2+\zeta_2]/4\pi+O(\varepsilon^3)$. With this counting the wave-function couplings feed back into the potential only at order $\varepsilon^3$, and at $d=4-\varepsilon$ the relevant beta functions are, up to $O(\varepsilon^3)$,
\begin{eqnarray}\label{eps_betas}
\beta_{\lambda_4}&=&-\varepsilon\lambda_4+\tfrac{N+8}{16\pi^2}\,\lambda_4^2\,,\nonumber\\
\beta_{\psi_2}&=&2\psi_2-\tfrac{2\lambda_4^2}{3(4\pi)^3}\,,\nonumber\\
\beta_{\zeta_2}&=&2\zeta_2-\tfrac{(N+8)\,\lambda_4^2}{3(4\pi)^3}\,.
\end{eqnarray}
The fixed point is thus at $\lambda_4^*=(4\pi)^2\varepsilon/(N+8)$, $\psi_2^*=\lambda_4^{*2}/[3(4\pi)^3]$ and $\zeta_2^*=(N+8)\lambda_4^{*2}/[6(4\pi)^3]$: the transverse coupling is independent of $N$, while the longitudinal one carries the same factor $N+8$ as the one-loop coefficient of $\beta_{\lambda_4}$. Substituting into \eqref{eta_rho0} we obtain
\begin{equation}
\eta=\tfrac{N+2}{2(N+8)^2}\,\varepsilon^2+O(\varepsilon^3)\,.
\label{eta_epsilon_result}
\end{equation}
This coincides with the standard two-loop $\varepsilon$-expansion result for the $O(N)$ universality class \cite{Zinn-Justin:2002ecy}, and reduces to the {\tt Ising} result $\eta=\varepsilon^2/54$ for $N=1$.

\subsection*{\color{teal}Large-$N$ limit}\vspace{-0.2cm}

In the limit $N\to\infty$ the FDR flow can be solved analytically. Rescaling $\rho\to N\rho$ and $u\to N u$, at leading order in $1/N$ only the $N-1$ transverse modes contribute to \eqref{FDR-LPA}, while the wave-function renormalizations do not feed back into the potential and $\eta=O(1/N)$. Setting $z_T=1$, the flow of $w(\rho)\equiv u'(\rho)$ reads
\begin{equation}
\partial_t w=-2w+\Big[(d-2)\rho-\tfrac{1}{4\pi}\,e^{-\frac{w}{4\pi}}\Big]w'\,.
\end{equation}
As usual at large $N$, the equation becomes linear when written for the inverse function $\rho(w)$ \cite{Tetradis:1995br,DAttanasio:1997he}:
\begin{equation}\label{largeNflow}
\partial_t\rho=2w\,\rho'-(d-2)\,\rho+\tfrac{1}{4\pi}\,e^{-\frac{w}{4\pi}}\,.
\end{equation}
For $2<d<4$, the Wilson--Fisher fixed point is the solution of \eqref{largeNflow} that is analytic at $w=0$,
\begin{equation}
\rho_*(w)=\frac{1}{4\pi}\sum_{k\geq0}\tfrac{(-1)^k}{k!\,\big(d-2-2k\big)}\Big(\frac{w}{4\pi}\Big)^k\,,
\end{equation}
so that the minimum of the potential sits at $\kappa_*=\rho_*(0)=1/[4\pi(d-2)]$, i.e.\ $\kappa_*\simeq N/(4\pi)$ in $d=3$ in the original variables.
Linear perturbations $\delta\rho\propto w^m e^{\theta t}$, analytic at $w=0$, have
\begin{equation}
\theta_m=2m-(d-2)\,,\qquad m=0,1,2,\dots\,,
\end{equation}
which gives $\nu=1/(d-2)$, $\omega=4-d$ and the whole tower of subleading exponents. These values do not depend on the form of the exponential threshold function: as for any functional RG at leading order in $1/N$ \cite{DAttanasio:1997he}, the FDR flow is exact in this limit. We checked these results numerically: at $N=2\cdot10^4$ the polynomial truncation gives $\kappa_*/N=0.079575$ and the eigenvalues $-1.000$, $1.000$, $3.000$, $5.000$ in $d=3$.

At next-to-leading order the wave-function renormalizations become important and the derivative expansion is no longer exact. Fitting our FDR-DE2 results for $100\leq N\leq1000$ we find
\begin{equation}
\eta\simeq\tfrac{0.314}{N}\,,\qquad
\nu\simeq1-\tfrac{1.13}{N}\,,\qquad
\omega\simeq1-\tfrac{2.93}{N}\,,
\end{equation}
to be compared with the exact coefficients $8/(3\pi^2)\simeq0.270$, $32/(3\pi^2)\simeq1.081$ and $64/(3\pi^2)\simeq2.161$ \cite{Moshe:2003xn}. The $1/N$ correction to $\nu$ is reproduced within $4\%$, while those to $\eta$ and $\omega$ are overestimated by about $16\%$ and $35\%$, respectively. This is the reason why the FDR curves in Figs.~\ref{figeta}--\ref{figomega} approach the large-$N$ ones only slowly, in particular for $\omega$.

\subsection*{\color{teal}Conclusion and Outlook}\vspace{-0.2cm}

In this work we applied the recently proposed Functional Dimensional Regularization (FDR) scheme to the $O(N)$ universality classes at second order in the derivative expansion. The method yields critical exponents with remarkable efficiency as a function of the number of fields $N$, reaching values that are competitive with higher-order Non-Perturbative RG (NPRG) and Conformal Bootstrap (CB) results. 
The predictions for $\eta$ are very competitive. For the universality classes {\tt Ising}, {\tt XY}, {\tt Heisenberg} and {\tt O(4)}, our predictions are comparable or even superior to NPRG at second order of the derivative expansion (DE2), and in some cases approach the raw fourth-order values. In the {\tt SAW} universality class, the FDR estimate of $\eta$ lies between the CB value \cite{Shimada:2015gda} and the values obtained from the six-loop $\varepsilon$E and from NPRG at fourth order of the derivative expansion (DE4).
A distinctive advantage of the FDR approach is its simplicity: the flow equations at DE2 fit on half a page and are known analytically, whereas the corresponding NPRG equations at DE4 typically span pages if written out explicitly and demand numerical interpolation. This drastic reduction in complexity translates into numerical stability and efficiency, allowing us to generate smooth curves of the critical exponents as functions of $N$ with minimal computational cost. These results demonstrate that FDR provides a competitive and efficient alternative for studying non-perturbative properties of scalar field theories.

Looking ahead, several promising directions arise. 
On the one hand, it is natural to push the method to higher orders, either by extending the derivative expansion to DE4 or by considering the two-loop terms in the main formula \eqref{masterformula}, to further test the robustness of the scheme. 
On the other hand, the $d$--dependence of the results remains to be explored in more detail, extending the present three-dimensional study to other values of $d$. 
Similarly, one can compute other universal quantities such as amplitude ratios \cite{DePolsi:2021cmi}  or study the Kosterlitz-Thouless transition in the $N=2$ model \cite{Grater:1994qx}.
A particularly intriguing aspect is the slight deterioration observed in $\nu$ at DE2 with respect to LPA, which deserves clarification and may lead to further refinements of the approach.
Finally, beyond the $O(N)$ case, the FDR framework is well suited to tackle more complex multi-component field theories, such as clock and Potts models, $O(N)\times O(2)$ systems, and the recently identified universality class with $N=4$ at $d_c=4$ \cite{Codello:2020lta}.
These avenues highlight the potential of FDR as a versatile and efficient framework to explore universality and critical phenomena across a wide range of models.

\subsection*{\color{teal}Acknowledgments}
\vspace{-0.2cm}

We thank C. A. S\'anchez-Villalobos and K. Falls for comments and feedback. 
The authors acknowledge financial and compute support from the CSIC grant I+D-2022-22520220100174UD.

\bibliography{referenceFDRon}

\onecolumngrid
\appendix

\section{\color{teal} Full, Strict and Strict/Light flows}\label{appC}

\begin{figure*}
\centering
\includegraphics[width=0.95\linewidth]{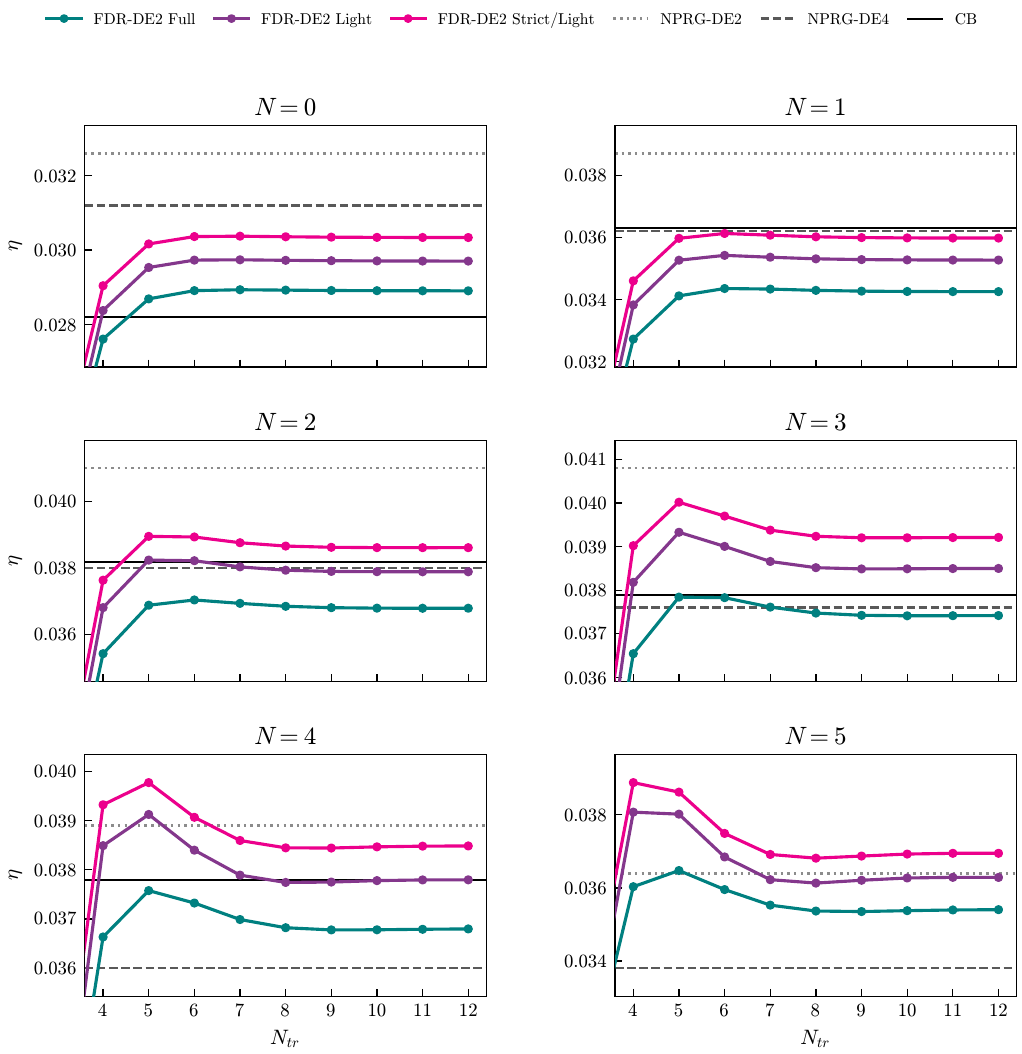}
\caption{Anomalous dimension $\eta$ in $d=3$ as a function of the truncation order $N_{tr}$, for $N=0,1,\dots,5$, obtained with the {\color{cfull}Full} (teal), {\color{clight}Light} (purple) and {\color{cstrict}Strict/Light} (magenta) FDR-DE2 flows defined in the text, expanding around $\rho=0$. The horizontal lines show the NPRG-DE2 (light grey, dotted) and NPRG-DE4 (dark grey, dashed) estimates \cite{DePolsi:2020pjk} and the CB value (black, solid) \cite{Shimada:2015gda,Kos:2016ysd,Chester:2019ifh,Kos:2015mba}; no CB value is shown for $N=5$.}
\label{fig:fullstrictlight}
\end{figure*}

The DE2 beta functionals \eqref{FDR-BetaU}, \eqref{betaT} and \eqref{betaL} contain terms that are formally of higher order in the wave-function renormalizations, whose treatment is not fixed by the derivative expansion itself. To assess their impact, we compare three versions of the flow, with the colour code of Fig.~\ref{fig:fullstrictlight}:
\begin{itemize}
\item {\color{cfull}Full}: the complete flow \eqref{FDR-BetaU}, \eqref{betaT}, \eqref{betaL} used in the main text.
\item {\color{clight}Light}: the wave-function renormalizations are set to one in the propagators of the one-loop diagrams that generate $\beta_{Z_T}$ and $\beta_{Z_L}$, i.e.\ $G_{T,L}(q)=1/(q^2+\omega_{T,L})$, while the vertices are left unchanged. The potential flow \eqref{FDR-BetaU} is kept as it is.
\item {\color{cstrict}Strict/Light}: in addition, all terms that are quadratic in the wave-function quantities $Z_L-Z_T$, $Z'_{T,L}$ and $Z''_{T,L}$ are dropped (``strict'' derivative expansion).
\end{itemize}
With $\Delta\equiv Z_L-Z_T=\rho Y$, $\omega_L=U'+2\rho U''$ and $\omega_L'=3U''+2\rho U'''$, the full beta functional \eqref{betaL} can be written as
\begin{eqnarray}
\beta_{Z_L}&=&\left\{-\frac{\mu^{d-2}}{4\pi}\,\frac{\textcolor{clight}{Z_L}\big(Z_L'+2\rho Z_L''\big)-\textcolor{cstrict}{3\rho Z_L'^2}}{\textcolor{clight}{Z_L^2}}+\frac{\mu^{d-4}}{(4\pi)^2}\,\frac{2\rho Z_L'\big(\textcolor{clight}{Z_L}\,\omega_L'-\textcolor{cstrict}{\omega_L Z_L'}\big)}{\textcolor{clight}{Z_L^3}}\right.\nonumber\\
&&\left.\hspace{0.6cm}-\frac{\mu^{d-6}}{(4\pi)^3}\,\frac{\rho}{3}\,\frac{\textcolor{clight}{Z_L^2}\,\omega_L'^2-2\textcolor{clight}{Z_L}\,\omega_L\omega_L'Z_L'+\textcolor{cstrict}{\omega_L^2Z_L'^2}}{\textcolor{clight}{Z_L^4}}\right\}e^{-\frac{\omega_L}{4\pi\mu^2\textcolor{clight}{Z_L}}}\nonumber\\
&&-(N-1)\left\{\frac{\mu^{d-6}}{(4\pi)^3}\,\frac{\rho}{3}\,\frac{U''^2\textcolor{clight}{Z_T^2}-2U'U''\textcolor{clight}{Z_T}\,Z_T'+\textcolor{cstrict}{U'^2Z_T'^2}}{\textcolor{clight}{Z_T^4}}+\frac{\mu^{d-4}}{(4\pi)^2}\,\frac{2\big(U''\textcolor{clight}{Z_T}-\textcolor{cstrict}{U'Z_T'}\big)\big(\rho Z_T'-\Delta\big)}{\textcolor{clight}{Z_T^3}}\right.\nonumber\\
&&\left.\hspace{1.6cm}+\frac{\mu^{d-2}}{4\pi}\,\frac{\textcolor{clight}{Z_T}\big(\rho Z_L'-\Delta\big)+\textcolor{cstrict}{\rho^2Z_T'^2-2\rho Z_T'\Delta}}{\rho\,\textcolor{clight}{Z_T^2}}\right\}e^{-\frac{U'}{4\pi\mu^2\textcolor{clight}{Z_T}}}\,.\label{betaLlight}
\end{eqnarray}
The wave-function renormalizations in purple are set to one in the {\color{clight}Light} approximation: they are the undifferentiated $Z_{T,L}$ coming from the propagators, while the difference $\Delta=\rho Y$ comes from the vertices and is kept. The terms in magenta are quadratic in the wave-function quantities $\Delta$, $Z'_{T,L}$ and $Z''_{T,L}$, and are dropped in the {\color{cstrict}Strict/Light} flow.

For $\beta_{Z_T}$ the Light approximation cannot be obtained by a substitution in \eqref{betaT}: there the denominator $\rho(Z_T-Z_L)U'+2\rho^2Z_TU''$, which originates from the difference of the transverse and longitudinal propagators, mixes the wave-functions of the propagators with the vertex quantity $\Delta$. Computing the diagrams with $G_{T,L}(q)=1/(q^2+\omega_{T,L})$ one finds instead
\begin{eqnarray}
\beta_{Z_T}^{\rm Light}&=&\left\{\mu^d\,\frac{1+2\rho Z_T'+\textcolor{cstrict}{\rho^2Z_T'^2}}{\rho^2U''}-\frac{\mu^{d-2}}{4\pi}\,\frac{1+(N-1)\rho Z_T'}{\rho}+\Delta\left[\frac{\mu^{d-2}}{4\pi}\,\frac{U'+\rho U''}{\rho^2U''}-\mu^d\,\frac{\omega_L\big(1+\textcolor{cstrict}{\rho Z_T'}\big)}{\rho^3U''^2}\right]\right.\nonumber\\
&&\left.\hspace{0.7cm}-\textcolor{cstrict}{\Delta^2\left[\frac{\mu^{d-2}}{16\pi}\,\frac{U'(U'+4\rho U'')}{\rho^3U''^2}-\mu^d\,\frac{\omega_L^2}{4\rho^4U''^3}\right]}\right\}e^{-\frac{U'}{4\pi\mu^2}}\nonumber\\
&&-\left\{\mu^d\,\frac{1+2\rho Z_T'+\textcolor{cstrict}{\rho^2Z_T'^2}}{\rho^2U''}+\frac{\mu^{d-2}}{4\pi}\,\frac{1+5\rho Z_T'+2\rho^2Z_T''}{\rho}-\Delta\,\omega_L\left[\frac{\mu^{d-2}}{4\pi}\,\frac{1+\textcolor{cstrict}{2\rho Z_T'}}{\rho^2U''}+\mu^d\,\frac{1+\textcolor{cstrict}{\rho Z_T'}}{\rho^3U''^2}\right]\right.\nonumber\\
&&\left.\hspace{0.7cm}+\textcolor{cstrict}{\Delta^2\omega_L^2\left[\frac{\mu^{d-2}}{16\pi\rho^3U''^2}+\frac{\mu^d}{4\rho^4U''^3}\right]}\right\}e^{-\frac{\omega_L}{4\pi\mu^2}}\,,\label{betaTlight}
\end{eqnarray}
where again the terms in magenta are dropped in the Strict/Light flow. Compared with \eqref{betaT}, the denominator becomes $2\rho^2U''$ and $\Delta$ appears at most quadratically.
Since both approximations break the invariance under rescalings of the field, their results depend on where the wave-functions are normalized. In Fig.~\ref{fig:fullstrictlight} we used the expansion around $\rho=0$ introduced in the section on the $\varepsilon$-expansion, with $z_L(0)=z_T(0)=1$, $N_{tr}$ potential couplings and $N_{tr}-2$ couplings in each wave-function.

Figure~\ref{fig:fullstrictlight} shows the resulting estimates of $\eta$ as a function of the truncation order $N_{tr}$ for $N=0,\dots,5$. All three versions converge rapidly, with plateaus reached for $N_{tr}\gtrsim 8$, and they are ordered in the same way for all $N$: Full gives the smallest value, Light an intermediate one and Strict/Light the largest. The spread between them is between $0.0014$ and $0.0018$ for all $N$, i.e.\ about $5\%$ of $\eta$.

For $N=1$, Strict/Light gives $\eta\simeq0.0360$, very close to the CB value $\eta=0.03630$ \cite{Kos:2016ysd}; this is the {\tt Ising} estimate of \cite{Beretta:2026zcy}. The comparison with the Full and Light results ($\eta\simeq0.0343$ and $0.0353$) shows that this agreement is partly fortuitous, since the different versions are equally consistent at order DE2. At the same time, the band spanned by the three versions provides an estimate of the uncertainty of the DE2 result, and it is informative: for $N=2,3,4$ the CB value lies inside the band, for $N=1$ it lies at its upper edge, while for $N=0$ the band lies between the CB value and the NPRG-DE4 one. The NPRG-DE4 values lie inside the band for $N=2,3$ and below it for $N=4,5$.

\end{document}